\documentclass[11pt,notitlepage,nofootinbib]{revtex4-1}

\usepackage{epsf}
\usepackage{latexsym}
\usepackage{amssymb}
\usepackage{amsmath}
\usepackage{bm}
\usepackage[dvips]{graphicx}
\usepackage{array}

\date{\today}
\def\d3{^{(3)}\nabla}

\begin{document}

\title{Effects on the CMB from magnetic field dissipation before recombination}

\author{Kerstin E. Kunze}

\email{kkunze@usal.es}

\affiliation{Departamento de F\'\i sica Fundamental and {\sl IUFFyM}, Universidad de Salamanca,
 Plaza de la Merced s/n, 37008 Salamanca, Spain}

\begin{abstract}
Magnetic fields present before decoupling are damped due to radiative viscosity. This energy injection affects the thermal and ionization history of the cosmic plasma. The implications for the CMB anisotropies and polarization are investigated for different parameter choices of a non helical stochastic magnetic field.
Assuming  a Gaussian smoothing scale determined by the magnetic damping wave number at recombination it is found that magnetic fields with present day strength less than  0.1 nG and negative magnetic spectral indices have a sizeable  effect on the CMB temperature anisotropies and polarization.
\end{abstract}

\maketitle

\section{Introduction}
\label{s0}
\setcounter{equation}{0}

Magnetic fields on large scales are redshifted with the expansion of the universe. Due to the non trivial interaction of the ionized component of the
baryon fluid with the rest of the cosmic plasma they suffer additional damping.
This dynamics depends on the physical conditions of the universe. Before recombination baryons and photons are tightly coupled. Thomson scattering is
very efficient at high temperatures and frequent scattering of photons off free electrons ensures the strong coupling of the baryon and photon fluids.
As the temperature falls Thomson scattering becomes less effective, the two fluids start decoupling introducing photon viscosity in the dynamics of the 
baryon fluid. Photons start diffusing and free streaming thereby dragging the baryons out of the gravitational potential wells.
This leads to a  damping of the density perturbation spectrum on scales below the photon diffusion scale, the Silk scale \cite{silk,kaiser}. 

Cosmic magnetic fields present before decoupling are tied to the completely ionized baryon fluid. Thus prior to decoupling radiative viscosity leads to damping of magnetic fields on small scales.
However, contrary to perturbations in a non magnetized plasma magnetic fields survive damping down to scales shorter than the Silk scale.
Perturbations in a  magnetized plasma comprise of three different modes which are the slow and fast magnetosonic modes and the Alfv\'en mode.
Whereas the first two are compressional the latter is not. As shown in \cite{jko,sb} slow magnetosonic modes and Alfv\'en modes enter an overdamped regime thereby surviving damping beyond the Silk scale. On the other hand, fast magnetosonic waves are damped at the Silk scale.

After decoupling radiative viscosity rapidly becomes suppressed and MHD turbulence can develop. Nonlinear interaction between different scales leads to 
decay of MHD turbulence and dissipation  of the magnetic field energy. Inverse cascade transfers energy from small to large scales. This has been observed in numerical simulations for helical magnetic fields and recently also for non helical ones \cite{KTBN,BKT,KBT}.
Even after recombination matter is not completely neutral, but rather there is a  small fraction of matter which remains ionized. 
This leads to plasma drift \footnote{This is often called ambipolar diffusion which, however, strictly speaking refers to diffusion of electrons and ions due to electrostatic coupling \cite{zweibel,mestel}. As pointed out in \cite{mestel} it is more appropriate to refer to it as plasma drift as the ions drift w.r.t. the plasma.}
caused by different velocities for the neutrals and ions in a magnetized plasma. The resulting friction force between the two components is of the order of the Lorentz term resulting in damping of the magnetic field.

The energy liberated by the damping of the cosmic magnetic field leads, on the one hand, to heating of matter, and, on the other hand, to spectral distortions of the photon spectrum. The resulting spectral distortions of the cosmic microwave background (CMB) have been calculated in the pre- \cite{JKO-dist,KuKo14}
as well as the  post-recombination \cite{SeSu, KuKo14,WB} era. In \cite{kuko15,CPF,planck15-pmf} the effects on the CMB temperature anisotropies and polarization due to the change in the thermal evolution in the post recombination universe have been studied. In this case the evolution of the matter temperature receives two additional source functions due to the dissipation of magnetic energy by decaying MHD turbulence as well as plasma drift. The former being important for large values of redshift close to decoupling and the latter being dominant in the more recent universe.

Here the effect of magnetic field dissipation in the pre-recombination universe on the CMB temperature anisotropies and polarization is considered.
In section \ref{s2} the corresponding angular power spectra are calculated considering the adiabatic, primordial curvature mode in the presence of the modified thermal and hence ionization history. It is known that magnetic fields also contribute to the final temperature anisotropies and polarization
(e.g. \cite{kb,pfp,sl,kk11}).
However, for the values of the magnetic field strengths considered here, these contributions can  be neglected as a first approximation.
In section \ref{s3} our conclusions are presented.

\section{Including magnetic field dissipation in the pre-recombination universe}
\label{s2}
\setcounter{equation}{0}

In  \cite{KuKo14} the induced CMB spectral distortions in the pre-recombination universe have been calculated and we follow that description here.
The magnetic field is assumed to be a non-helical, Gaussian random field determined by its two-point function in Fourier space given by
\begin{eqnarray}
\langle B_i^*(\vec{k})B_j(\vec{q})\rangle=(2\pi)^3\delta({\vec{k}-\vec{q}})P_B(k)\left(\delta_{ij}-\frac{k_ik_j}{k^2}\right),
\end{eqnarray}
where the power spectrum, $P_B(k)$, is assumed to be a power law,
$P_B(k)=A_Bk^{n_B}$, with the amplitude, $A_B$, and the spectral index, $n_B$.
The ensemble average energy density of the magnetic field is defined using a Gaussian window function so that
\begin{eqnarray}
\langle\rho_{B,0}\rangle=\int\frac{d^3k}{(2\pi)^3}P_{B,0}(k)e^{-2\left(\frac{k}{k_c}\right)^2}, 
\end{eqnarray}
where $k_c$ is a certain Gaussian smoothing scale and a "0" refers to the present epoch. 
It is convenient to express the magnetic field power spectrum in terms of $\rho_{B,0}$ yielding
\begin{eqnarray}
P_{B,0}(k)=\frac{4\pi^2}{k_c^3}\frac{2^{(n_B+3)/2}}{\Gamma\left(\frac{n_B+3}{2}\right)}\left(\frac{k}{k_c}\right)^{n_B}\langle\rho_{B,0}\rangle.
\end{eqnarray}
$n_B=3$ corresponds to the scale-invariant case 
for which the contribution to the energy density per logarithmic wavenumber
is independent of wave number.
The spectral index depends on the details of the generation mechanism of the magnetic field.
Whereas inflationary produced magnetic fields generally have negative spectral indices \cite{tw} those 
generated by causal processes such as during the electroweak phase transition \cite{ew1,ew2,ew3, rev1,rev2,rev3,rev4}
have positive values. Moreover it was shown in \cite{dc} that in the latter case they have to be 
an even integer with $n_B\geq 2$.

\begin{figure}[h!]
\centerline{\epsfxsize=3.5in\epsfbox{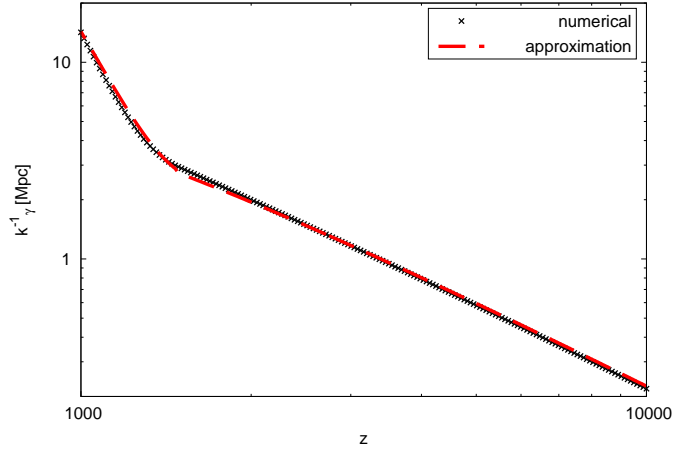}}
\caption{Numerical and approximate solution for the photon diffusion scale. Cosmological parameters are set to the best fit values of the WMAP 9 year only data \cite{wmap9}.}
\label{fig1}
\end{figure}

\begin{figure}[h!]
\centerline{\epsfxsize=3.2in\epsfbox{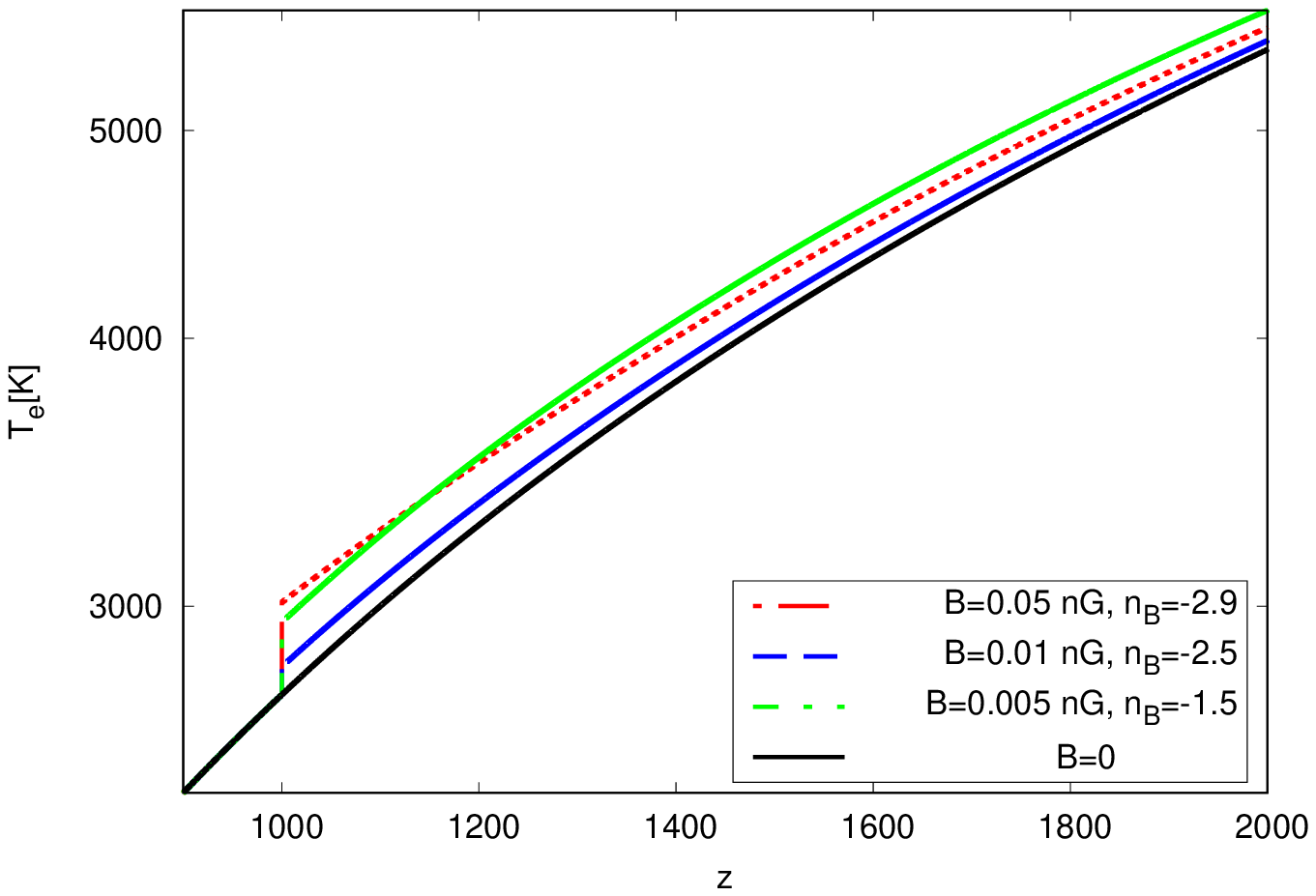}\hspace{0.5cm}
\epsfxsize=3.2in\epsfbox{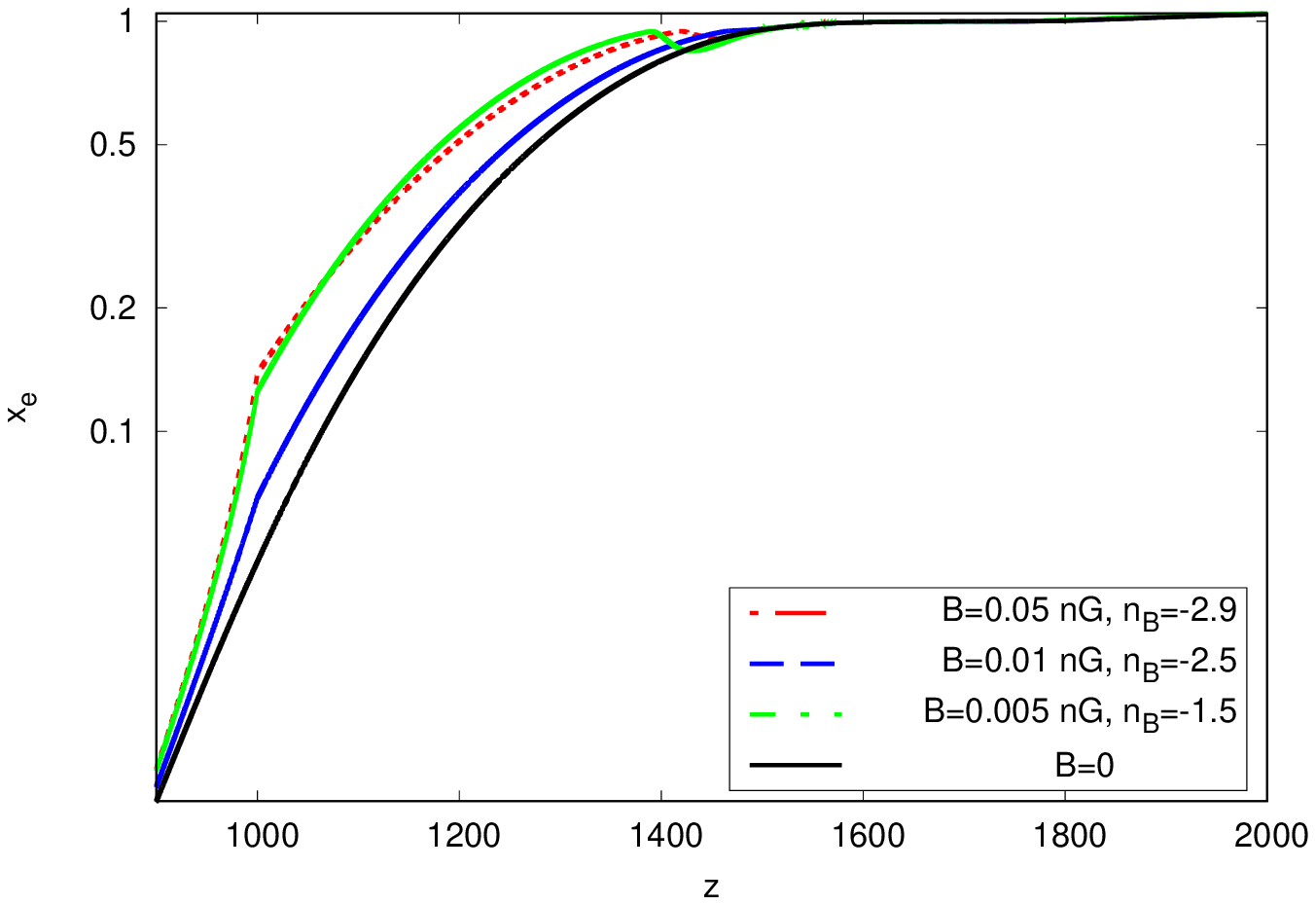}}
\centerline{\epsfxsize=3.2in\epsfbox{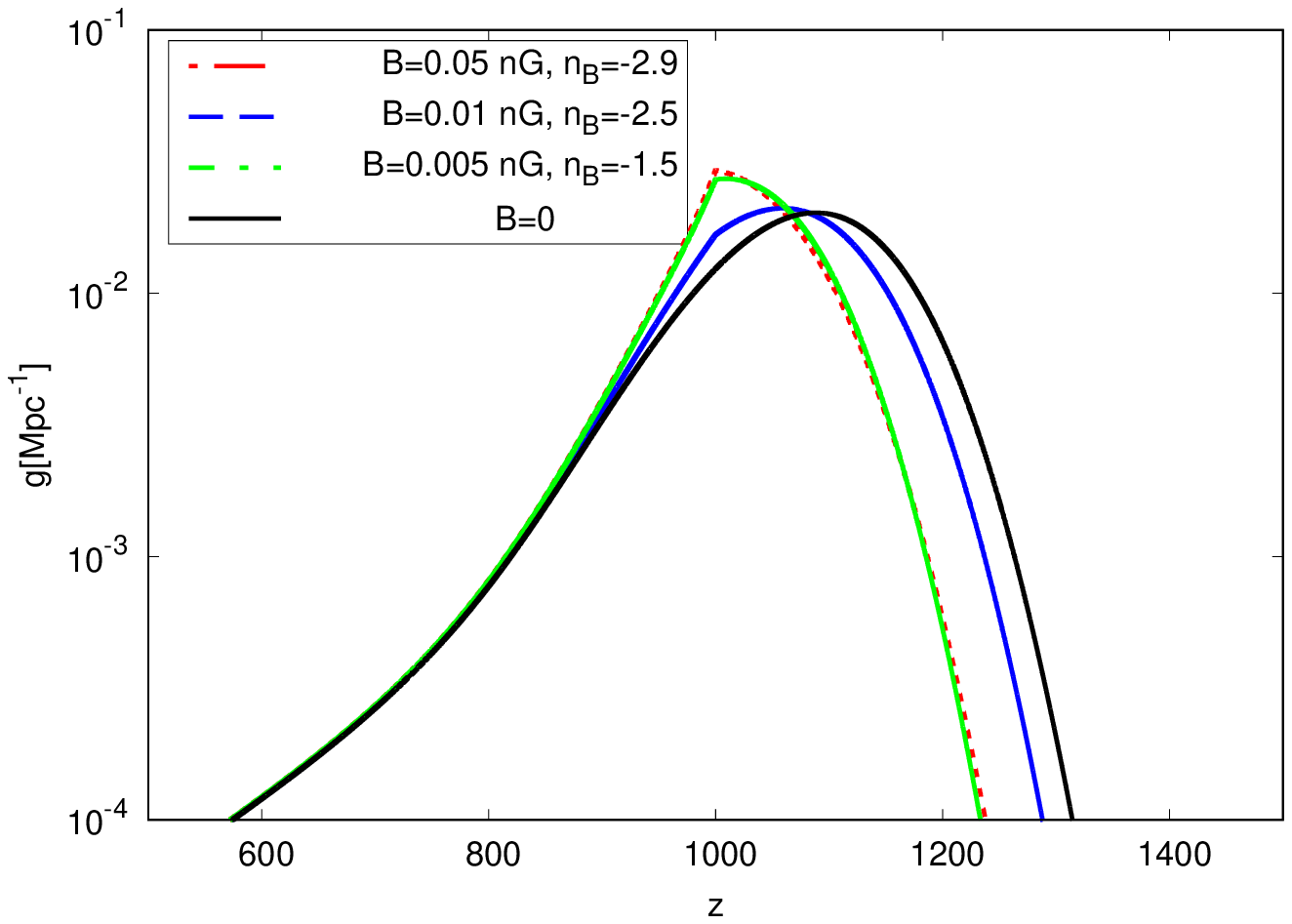}\hspace{0.5cm}
\epsfxsize=3.2in\epsfbox{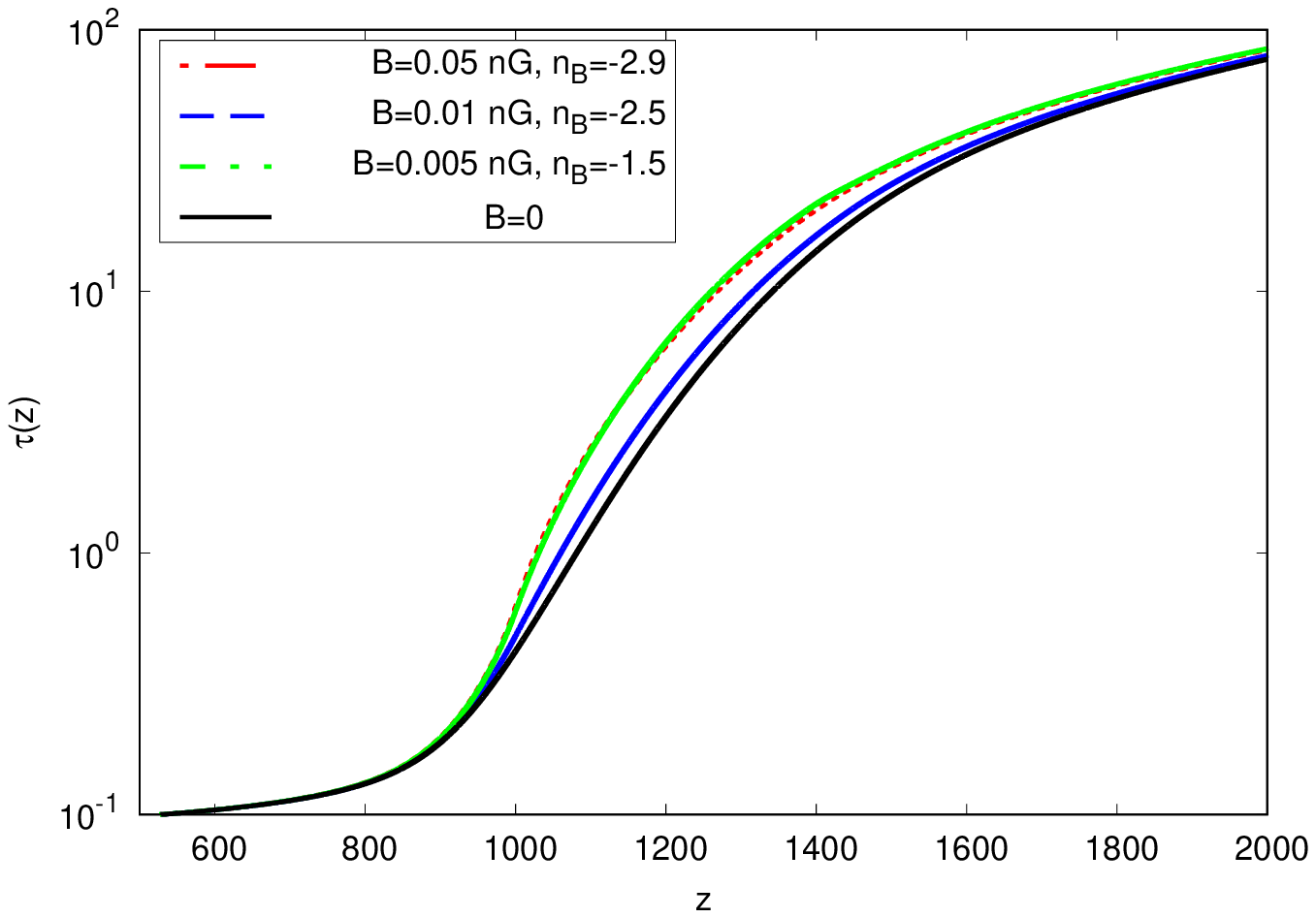}
}
\caption{Thermal and ionization history for different choices of the magnetic field parameters, shown are 
the electron temperature $T_e$ ({\it top left}), the ionization fraction $x_e$ ({\it top right}), the visibility function $g$ ({\it bottom left}) and the optical depth $\tau$ ({\it bottom right}). The Gaussian smoothing scale is determined by  the maximal damping wave number at decoupling hence $k_c=k_{d,dec}$.
Cosmological parameters are set to the best fit values of the WMAP 9 year only data \cite{wmap9-http}.}
\label{fig2}
\end{figure}

The volume energy injection rate  due to damping of magnetosonic and Alfv\'en waves is determined by
$\dot{Q}=a^{-4}d(\rho_B a^4)/dt$ where the comoving magnetic energy density is calculated setting the 
smoothing scale to be the damping scale $k_d$ leading to \cite{KuKo14}
\begin{eqnarray}
\nonumber
a^4\langle
\rho_{B}\rangle(z)&=&a^4\int\frac{d^3k}{(2\pi)^3}P_B(k)e^{-2\left(\frac{k}{k_d(z)}\right)^2}\\
&=&
 a^4_0\int\frac{d^3k}{(2\pi)^3}P_{B,0}(k)e^{-2\left(\frac{k}{k_d(z)}\right)^2}.
\end{eqnarray}
This yields the volume energy injection rate 
\begin{eqnarray}
\frac{dQ}{dz}=-\frac{n_B+3}{2}\rho_{\gamma,0}(1+z)^4
\left(\frac{\rho_{B,0}}{\rho_{\gamma,0}}\right)k_c^{-(n_B+3)}k_d(z)^{n_B+5}\frac{d}{dz}k_d^{-2}(z), 
\label{dQodz}
\end{eqnarray}
where $\frac{\rho_{B,0}}{\rho_{\gamma,0}}=9.545\times 10^{-8}\left(\frac{B_0}{\rm nG}\right)^2$ for $T_{\rm CMB}=2.725$ K.
The damping wave number of the magnetic field is given in terms of the photon diffusion wave number $k_{\gamma}$
yielding \cite{sb, jko}
\begin{eqnarray}
k_d=\alpha k_{\gamma}
\label{kd} 
\end{eqnarray}
where $\alpha$ is given by
\begin{eqnarray}
\alpha=\left\{
\begin{array}{lr}
1 & {\rm fast\;\; ms \;\; modes}\\
(v_A\cos\theta)^{-1}\simeq \frac{2.6\times 10^3}{\cos\theta}(1 {\rm nG}/B_0)& 
{\rm \hspace{1cm} slow \;\; ms \;\; \& \;\;  A \;\; modes}
\end{array}
\right.
 \label{alpha}
 \end{eqnarray}
for fast, slow magnetosonic (ms) modes and Alfv\' en (A) modes. In the following the angle between the wave vector and the magnetic field vector, $\theta$, will be ignored, hence $\cos\theta=1$.
Including polarization the photon diffusion scale is given by \cite{kaiser}
\begin{eqnarray}
k_{\gamma}^{-2}(z)=\int_z^{\infty}\frac{dz}{6H(z)(1+R)\dot{\tau}}\left(\frac{16}{15}+\frac{R^2}{1+R}\right).
\label{exact}
\end{eqnarray}
The baryon-to-photon density ratio, $R$, is given by
 $R=\frac{3}{4}\frac{\rho_{b}}{\rho_{\gamma}}$ and $\dot{\tau}$ is the differential optical depth. 
The photon diffusion scale upto recombination can be approximated in terms of the hypergeometric function
$F(\alpha,\beta;\gamma;z)$  (for more details see the appendix \ref{app}).
The explicit form depends on the value of the redshift, i.e. if it is bigger or smaller than some redshift $z_*$ at which the 
approximate form of the differential optical redshift changes. 
For the WMAP 9 best-fit parameters
$z_{*}\simeq 1486.57$   \cite{wmap9, KuKo14}.
For  $z>z_{*}$ it is given by
\begin{eqnarray}
k_{\gamma, \; z\geq z_*}^{-2}(z)&=&2.16567\times 10^7\left(\Omega_{r,0}h^2\right)^{-\frac{1}{2}}\left(1-\frac{Y_p}{2}\right)^{-1}\left(\Omega_b h^2\right)^{-1}
\nonumber\\
&\times&
\frac{16}{15}\left[\frac{1}{3z^3} F\left(\frac{1}{2},3;4;-\frac{2+z_{eq}}{z}\right)-\frac{0.1875}{z^4}\frac{\Omega_{b,0}}{\Omega_{\gamma,0}}
F\left(\frac{1}{2},4;5;-\frac{2+z_{eq}}{z}\right)\right] \;\; {\rm Mpc}^2
\label{kga1}
\end{eqnarray}
and for $z_{\rm dec}<z<z_*$  by
\begin{eqnarray}
 k_{\gamma, \;z<z_*}^{-2}(z)&=&0.05 k_{\gamma, \; z\geq z_*}^{-2}(z_*)+1.49402\times 10^6(\Omega_{r,0}h^2)^{-1}\Omega_b^{-c_1}c_2^{-1}10^{3c_2}
 \frac{8}{15(2+c_2)(3+c_2)\Omega_{\gamma,0}}
 \nonumber\\
&\times&\left[z^{-3-c_2}\left[4\Omega_{\gamma,0}(3+c_2)zF\left(1,2+c_2;3+c_2;-\frac{2+z_{eq}}{z}\right)
\right.\right.
\nonumber\\
&&\left.\left.
-3(2+c_2)\Omega_{b,0}F\left(1,3+c_2;4+c_2;-\frac{2+z_{eq}}{z}\right)\right]
\right.
\nonumber\\
&&\left.
-z_*^{-3-c_2}\left[4\Omega_{\gamma,0}(3+c_2)z_*F\left(1,2+c_2;3+c_2;-\frac{2+z_{eq}}{z_*}\right)
\right.\right.
\nonumber\\
&&\left.\left.
-3(2+c_2)\Omega_{b,0}F\left(1,3+c_2;4+c_2;-\frac{2+z_{eq}}{z_*}\right)\right]
\right] \;\; {\rm Mpc}^2
\label{kga2}
\end{eqnarray}
The approximate solution together with the numerical solution is shown in figure \ref{fig1}.
Assuming equipartition of energy in the three magnetic modes the matter heating rate receives a factor $\frac{2}{3}$ if only the dominant part due to the dissipation of slow magnetosonic and Alfv\' en modes is considered \cite{KuKo14}. Moreover, there is an additional factor of $\frac{2}{3}$ taking into account that not all of the injected energy goes into heating but rather $\frac{1}{3}$ of it causes spectral distortions \cite{cks,ksc2,pz}. 
Thus the electron temperature evolves as
\begin{eqnarray}
\dot{T}_e=-2\frac{\dot{a}}{a}T_e+\frac{x_e}{1+x_e}\frac{8\rho_{\gamma}\sigma_T}{3m_ec}\left(T_{\gamma}-T_e\right)+\frac{x_e\Gamma}{1.5 k_B n_e},
\end{eqnarray}
where the heating rate $\Gamma=\frac{4}{9}\dot{Q}$ is calculated including the contributions from slow magnetosonic and Alfv\'en modes. Using in 
equation (\ref{dQodz}) the corresponding expression for the damping scale from equations  (\ref{kd}) and (\ref{alpha}) together with 
(\ref{kga1}) and (\ref{kga2}) allows to determine the effect of magnetic field dissipation before recombination on the thermal and ionization history as well as the angular power spectra of the CMB anisotropies of the temperature and polarization.
The Gaussian smoothing scale is set to the magnetic damping wave number at decoupling $k_c=k_{d,dec}$ where 
$k_{d,dec}=\frac{286.91}{\cos\theta}\left(\frac{\rm nG}{B_0}\right)$ Mpc$^{-1}$ for the  best-fit $\Lambda$CDM model for the WMAP 9-year data only \cite{wmap9,KuKo14}. This corresponds to the maximal value of the magnetic damping wave number.
For the numerical calculation the {\tt CLASS} code \cite{class1,class2,class3,class4,class5} has been adapted accordingly.
The results  are reported for different choices of the magnetic field parameters in figure \ref{fig2}.
The heating due to the magnetic field dissipation leads to a shift of the maximum in the visibility function to lower values of $z$. 
This implies a delay in recombination. The effect is stronger for spectral indices of the magnetic field far from the nearly scale invariant case, $n_B=-2.9$.
The width of the curve of the visibility function becomes narrower for smaller values of $n_B$ and peaks at a larger value. 
In figure {\ref{fig3} the effect on the angular power spectra of the auto- and cross correlation functions of the temperature (T) and polarization E-mode 
are shown for the same choice of magnetic field parameters as in figure \ref{fig2}.
The reduced width of the surface of last scattering leads to a shift of the peaks (cf. figure \ref{fig3}).
\begin{figure}[h!]
\centerline{\epsfxsize=3.2in\epsfbox{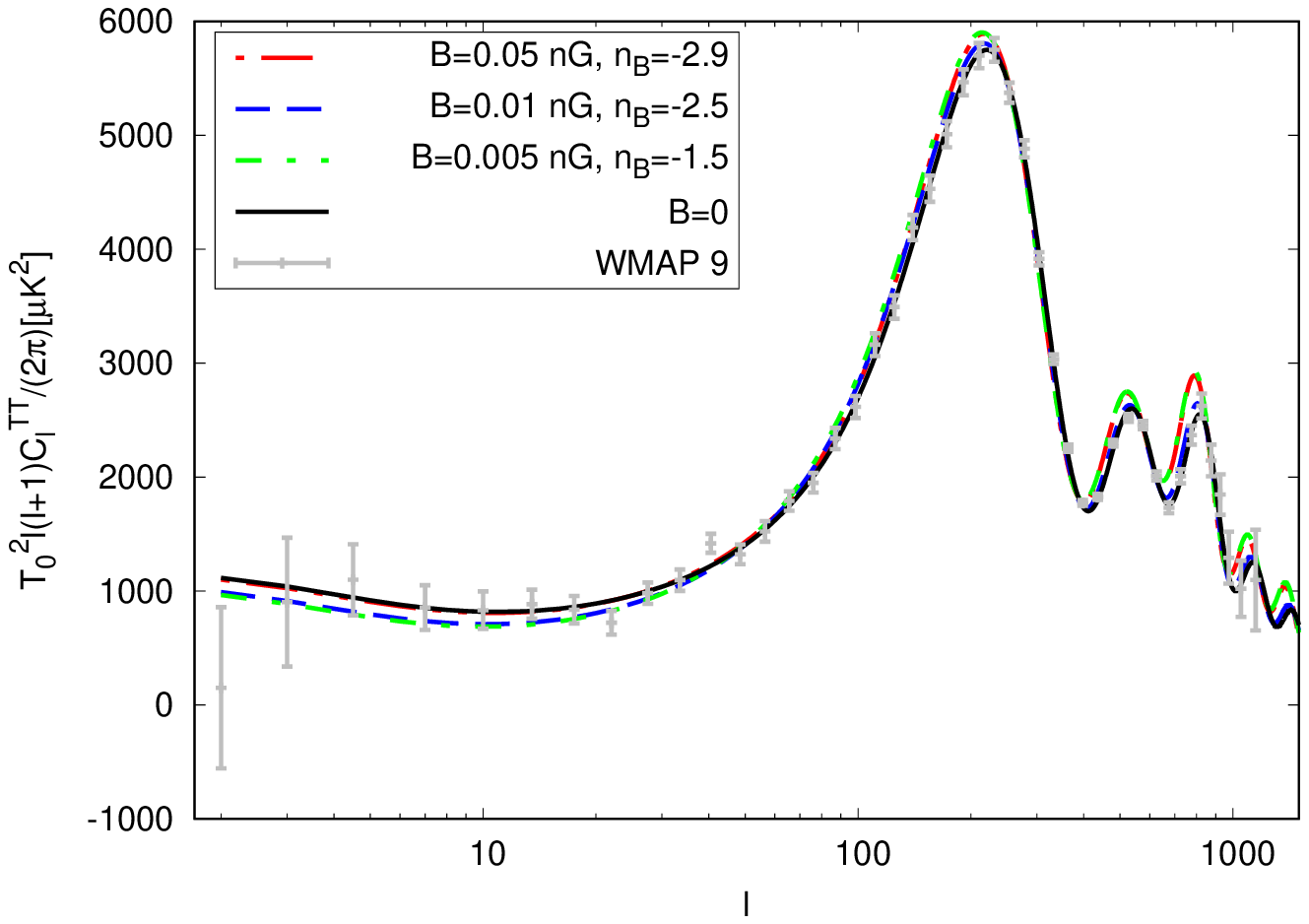}\hspace{0.5cm}
\epsfxsize=3.2in\epsfbox{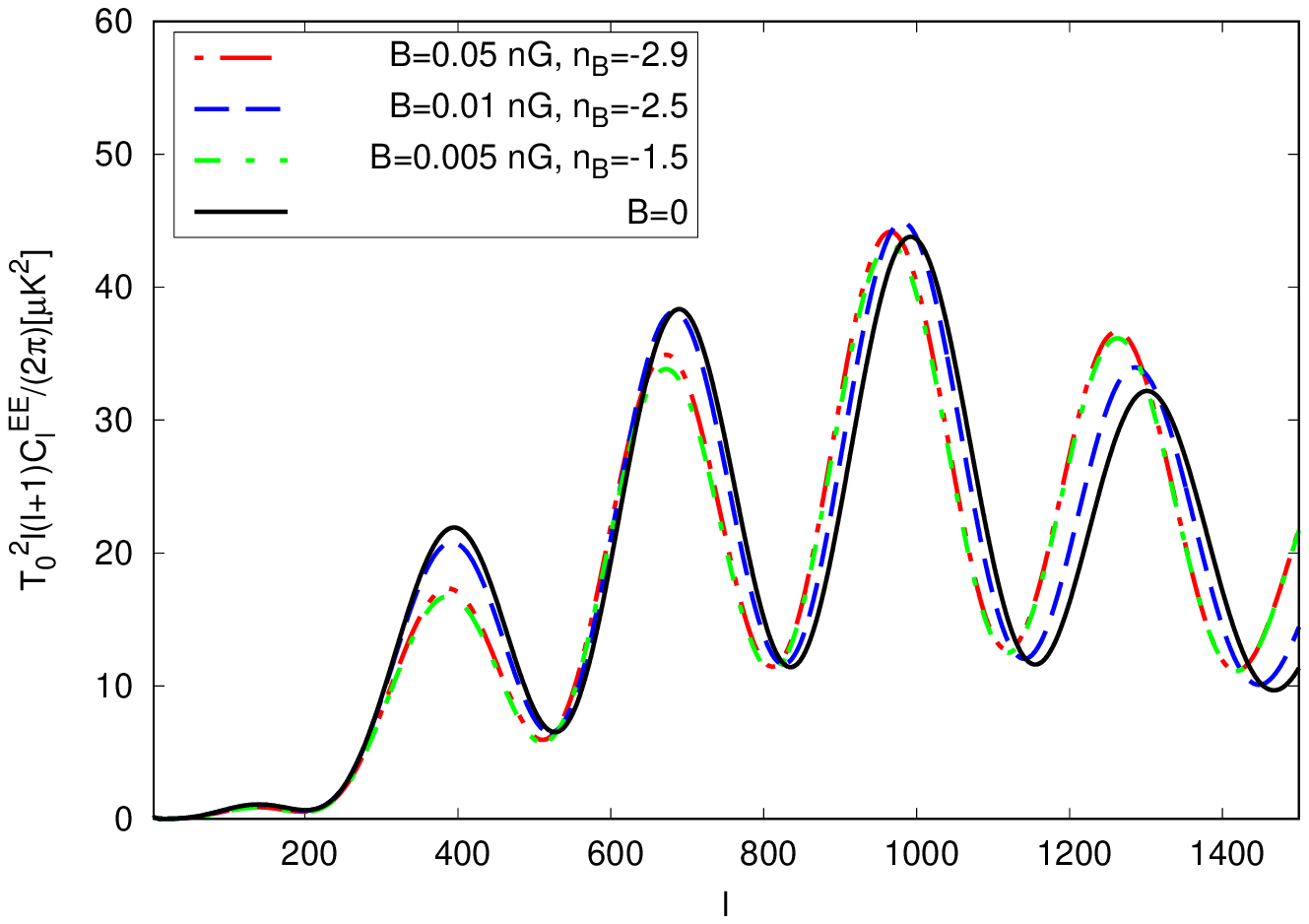}}
\centerline{\epsfxsize=3.2in\epsfbox{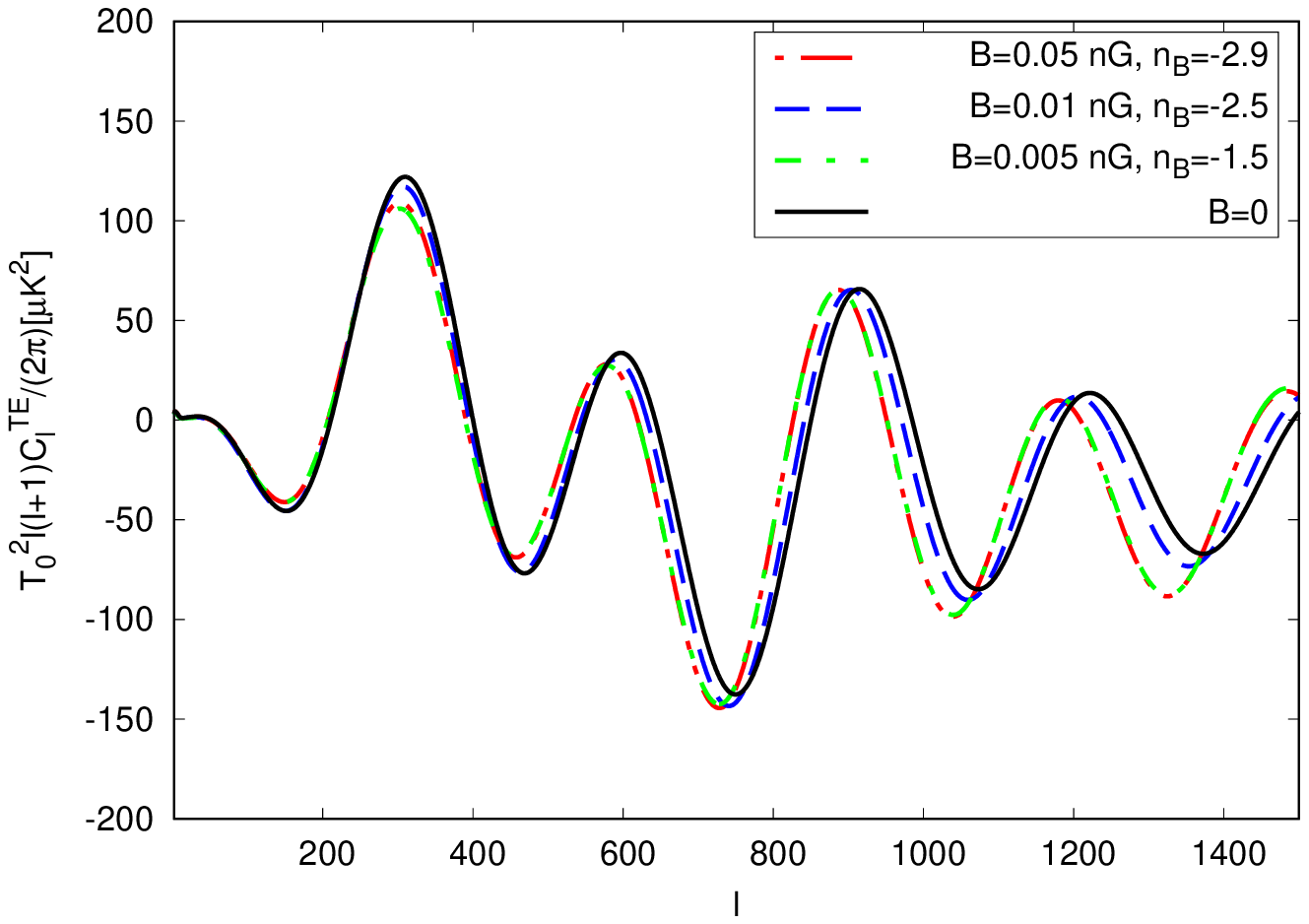}
}
\caption{Angular power spectrum of auto- and cross correlation functions of temperature (T) anisotropies and polarization E-mode for different choices of the magnetic field parameters. The Gaussian smoothing scale is determined by  the maximal damping wave number at decoupling, namely,  $k_c=k_{d,dec}$. Cosmological parameters are set to the best fit values of the WMAP 9 year only data \cite{wmap9}. For the temperature autocorrelation function the 9-year data of WMAP are included  \cite{wmap9-http}.}
\label{fig3}
\end{figure}
The WMAP 9 year temperature data (cf. figure \ref{fig3} ({\sl upper left panel})) indicate that for the chosen spectral indices the magnetic field amplitude has to be constrained to be less than of order of $10^{-3}$ nG. However, the precise values can only be determined with a full numerical parameter estimation which is currently under investigation \cite{kk-mc}.

\section{Conclusions }
\label{s3}
\setcounter{equation}{0}

We have investigated the effect of magnetic field dissipation before recombination on the thermal and ionization history of the universe. This affects the CMB angular power spectrum of temperature anisotropies and polarization. The magnetic field dissipation before recombination is complementary to the decaying MHD turbulence and plasma drift which take place in the nearly neutral, post recombination universe.
The dissipation of magnetic fields before recombination also leads to spectral distortions of the CMB. For redshifts between $2\times10^6$ and $5\times10^4$ this leads to $\mu$ distortion. In \cite{KuKo14} it was found that the COBE/FIRAS limit on $|\mu|<9\times 10^{-5}$ \cite{firas3} for the choice of Gaussian smoothing scale used here constrains the magnetic field amplitude to be $B_0<1.6$ nG for $n_B=-2.9$, $B_0<1.0$ nG for $n_B=-2.5$ and $B_0<0.1$ nG for $n_B=-1.5$. The projected PIXIE limit  $|\mu|<9\times 10^{-8}$ \cite{pixie} results in stronger constraints, namely, $B_0<1.0$ nG for $n_B=-2.9$, $B_0<0.1$ nG for $n_B=-2.5$ and $B_0<10^{-3}$ nG for $n_B=-1.5$. For the latter the constraint on the magnetic field amplitude is comparable 
to what is imposed by the observed spectrum of CMB temperature anisotropies (cf. figure \ref{fig3} ({\sl upper left panel})). However, in the other two cases these constraints are less stringent by several orders of magnitude.
For the choice of parameters used here the magnetic field parameters are quite strongly constrained when comparing with the WMAP 9 year temperature data \cite{kk-mc}. These constraints are stronger than from the corresponding constraint on the $\mu$ type spectral distortion for the projected PIXIE limits. This is similar to what is found in the case of the constraints on dark matter annihilation from the CMB and from spectral distortions from COBE/FIRAS \cite{fgls}.
For redshifts below $5\times 10^4$ spectral distortions  generated by energy injection  lead to $y$-type distortions. However, since there are several sources in the post recombination universe of $y$-type distortions it is more difficult to disentangle those from magnetic field dissipation.

\section{Acknowledgements}
Financial support by Spanish Science Ministry grants FPA2015-64041-C2-2-P (FEDER) 
is gratefully acknowledged. 
We acknowledge the use of the Legacy Archive for Microwave Background Data Analysis (LAMBDA), part of the High Energy Astrophysics Science Archive Center (HEASARC). HEASARC/LAMBDA is a service of the Astrophysics Science Division at the NASA Goddard Space Flight Center.

\appendix
\section{Approximation of the photon diffusion scale}
\label{app}
\numberwithin{equation}{section}
\setcounter{equation}{0}

The photon diffusion scale is given by \cite{kaiser}
\begin{eqnarray}
k_{\gamma}^{-2}(z)=\int_z^{\infty}\frac{dz}{6H(z)(1+R)\dot{\tau}}\left(\frac{16}{15}+\frac{R^2}{1+R}\right)
\label{exact}
\end{eqnarray} 
where
 $R=\frac{3}{4}\frac{\Omega_{b,0}}{\Omega_{\gamma,0}}(1+z)^{-1}$ is the baryon-to-photon density ratio.
Following \cite{HuSu} the differential optical depth can be approximated by
\begin{eqnarray}
\dot{\tau}(z)=\frac{c_2}{1000}\Omega_b^{c_1}\left(\frac{z}{1000}\right)^{c_2-1}\frac{\dot{a}}{a}\left(1+z\right),
\label{app1}
\end{eqnarray}
where a dot indicates the derivatives w.r.t. conformal time, $c_1=0.43$,
and $c_2=16+1.8\ln\Omega_b$. 
 The ionization fraction using $x_e(z)={\rm
min}(\dot{\tau}(n_e\sigma_T\frac{a}{a_0})^{-1},1)$, where
\begin{eqnarray}
\left(n_e(z)\sigma_T\frac{a}{a_0}\right)^{-1}=4.34\times 10^4\left(1-\frac{Y_p}{2}\right)^{-1}\left(\Omega_bh^2\right)^{-1}\left(\frac{T}{2.725 {\rm K}}\right)^{-3}\left(1+z\right)^{-2}\;\; {\rm Mpc}.
\end{eqnarray}
For the best-fit parameters of the {\sl WMAP} 9-year data only \cite{wmap9}, $x_e$
calculated using the expression (\ref{app1}) is larger than one for
$z>z_{*}\simeq 1486.57$ \cite{KuKo14}. Thus, the differential optical depth is given
by eq. (\ref{app1}) for $z_{\rm dec}<z<z_*$ and by
$\dot{\tau}=n_e\sigma_T\frac{a}{a_0}$ for $z\geq z_*$. Thus in the two regimes the photon diffusion scale is given by 
\begin{eqnarray}
k_{\gamma, \; z\geq z_*}^{-2}(z)&=&2.16567\times 10^7\left(\Omega_{r,0}h^2\right)^{-\frac{1}{2}}\left(1-\frac{Y_p}{2}\right)^{-1}\left(\Omega_b h^2\right)^{-1}
\nonumber\\
&\times&
\int_z^{\infty} dz (1+z)^{-\frac{7}{2}}(2+z+z_{eq})^{-\frac{1}{2}}(1+R)^{-1}\left(\frac{16}{15}+\frac{R^2}{1+R}\right) \;\; {\rm Mpc}^2
\label{numk.z.gt.z*}
\end{eqnarray}
and in $z_{\rm dec}<z<z_*$,
\begin{eqnarray}
k_{\gamma, \;z<z_*}^{-2}(z)&=&1.49402\times 10^6(\Omega_{r,0}h^2)^{-1}\Omega_b^{-c_1}c_2^{-1}10^{3c_2}
\nonumber\\
&\times&\int_{z}^{z_*}dz(1+z)^{-3}(2+z+z_{eq})^{-1}z^{1-c_2}(1+R)^{-1}\left(\frac{16}{15}+\frac{R^2}{1+R}\right) \;\; {\rm Mpc}^2
\nonumber\\
&+&k_{\gamma, \;z\geq z_*}^{-2}(z_*).
\label{numk.z.lt.z*}
\end{eqnarray}
Here the expansion rate $H(z)$ was used, 
$H(z)=H_0\Omega_{r,0}^{\frac{1}{2}}\left(1+z\right)^2\left(1+\frac{1+z_{eq}}{1+z}\right)^{\frac{1}{2}}$ where 
$\Omega_{r,0}=1.69\, \Omega_{\gamma,0}$ is the present-day total density
 of relativistic species including the 
standard value for the  effective number of light neutrinos,
 $N_\nu=3.04$. Moreover the epoch of radiation-matter equality is given by
 $\Omega_{r,0}=\Omega_{m,0}/(1+z_{eq})$. 
These expressions can be approximated by, for  $z>z_{*}\simeq 1486.57$
\begin{eqnarray}
k_{\gamma, \; z\geq z_*}^{-2}(z)&=&2.16567\times 10^7\left(\Omega_{r,0}h^2\right)^{-\frac{1}{2}}\left(1-\frac{Y_p}{2}\right)^{-1}\left(\Omega_b h^2\right)^{-1}
\nonumber\\
&\times&
\frac{16}{15}\left[\frac{1}{3z^3} F\left(\frac{1}{2},3;4;-\frac{2+z_{eq}}{z}\right)-\frac{0.1875}{z^4}\frac{\Omega_{b,0}}{\Omega_{\gamma,0}}
F\left(\frac{1}{2},4;5;-\frac{2+z_{eq}}{z}\right)\right] \;\; {\rm Mpc}^2
\end{eqnarray}
where $F(\alpha,\beta;\gamma,z)$ is the hypergeometric function \cite{GR}, and
 for $z_{\rm dec}<z<z_*$ by
\begin{eqnarray}
 k_{\gamma, \;z<z_*}^{-2}(z)&=&0.05 k_{\gamma, \; z\geq z_*}^{-2}(z_*)+1.49402\times 10^6(\Omega_{r,0}h^2)^{-1}\Omega_b^{-c_1}c_2^{-1}10^{3c_2}
 \frac{8}{15(2+c_2)(3+c_2)\Omega_{\gamma,0}}
 \nonumber\\
&\times&\left[z^{-3-c_2}\left[4\Omega_{\gamma,0}(3+c_2)zF\left(1,2+c_2;3+c_2;-\frac{2+z_{eq}}{z}\right)
\right.\right.
\nonumber\\
&&\left.\left.
-3(2+c_2)\Omega_{b,0}F\left(1,3+c_2;4+c_2;-\frac{2+z_{eq}}{z}\right)\right]
\right.
\nonumber\\
&&\left.
-z_*^{-3-c_2}\left[4\Omega_{\gamma,0}(3+c_2)z_*F\left(1,2+c_2;3+c_2;-\frac{2+z_{eq}}{z_*}\right)
\right.\right.
\nonumber\\
&&\left.\left.
-3(2+c_2)\Omega_{b,0}F\left(1,3+c_2;4+c_2;-\frac{2+z_{eq}}{z_*}\right)\right]
\right] \;\; {\rm Mpc}^2
\end{eqnarray}


\bibliography{references}

\bibliographystyle{apsrev}

\end{document}